\documentclass[runningheads]{llncs}
\usepackage[T1]{fontenc}
\usepackage{graphicx,verbatim}
\usepackage{placeins,float}
\usepackage{amsmath,amssymb,booktabs,xcolor}
\usepackage[normalem]{ulem}
\newcommand{\ci}[2]{\,{\scriptsize[#1, #2]}}
\usepackage[hidelinks]{hyperref}
\makeatletter
\let\prismOriginalFnsymbol\@fnsymbol
\def\@fnsymbol#1{\ifnum#1=2\ensuremath{\dagger}\else\prismOriginalFnsymbol{#1}\fi}
\makeatother

\begin{document}
\title{CryoDINO: A generalizable 3D foundation model for cryo-electron tomography}
%
\authorrunning{F. Author et al.}


\authorrunning{Anonymized Author et al.}
\author{
Ahmadreza Attarpour\inst{1}\thanks{Equal contribution.} \and
Sumin Kim\inst{2,3}\protect\footnotemark[1] \and
Tony Xu\inst{4,5} \and
Charles Chen\inst{1} \and
Daphne Baron\inst{1} \and
Ching-Yuan Yu\inst{1} \and
Yeonwoo Seo\inst{6} \and
Baidu Li\inst{10} \and
Elnaz Khalili Samani\inst{7} \and
Mohammad T. Mazhab-Jafari\inst{4,7} \and
Maged Goubran\inst{4,5,8,9} \and
Barry Rubin\inst{1} \and
Jun Ma\inst{4,7,10} \and
Bo Wang\inst{2,3,4,11}\thanks{Correspondence to: Ahmadreza Attarpour <ahmadreza.attarpour@uhn.ca>, Sumin Kim <sumin.kim@uhn.ca>, Bo Wang <bowang.wang@utoronto.ca>}
}

\titlerunning{CryoDINO}
\authorrunning{A. Attarpour and S. Kim et al.}

\institute{
Peter Munk Cardiac Centre, University Health Network, Toronto, Canada \and
Department of Computer Science, University of Toronto, Toronto, Canada \and
Vector Institute, Toronto, Canada \and
Department of Medical Biophysics, University of Toronto, Toronto, Canada \and
Physical Sciences Platform, Sunnybrook Research Institute, Toronto, Canada \and
Graduate School of AI, KAIST, Seoul, South Korea \and
Princess Margaret Cancer Center, University Health Network, Toronto, Ontario, Canada \and
Hurvitz Brain Sciences, Sunnybrook Health Sciences Centre, Toronto, Canada \and
Centre for Neuromodulation, Sunnybrook Health Sciences Centre, Toronto, Canada \and
AI Hub, University Health Network, Toronto, Ontario, Canada \and
Department of Laboratory Medicine and Pathobiology, University of Toronto, Toronto, Ontario, Canada
}

\maketitle
\begin{abstract}
Nanometre-resolution images acquired with cryogenic electron tomography (cryo-ET) provide unprecedented insights into sub-cellular architecture and the 3D spatial organization of macromolecules in their native environment. Yet, there remains an unmet need for computational pipelines that generalize across experimental conditions and downstream tasks while remaining robust to extremely low signal-to-noise ratios and reconstruction artifacts. We present CryoDINO, a general-purpose foundation model for cryo-ET, pretrained through self-supervised teacher–student distillation on an exceptionally large and diverse corpus of 1{,}658 tomograms and 425{,}852 3D patches spanning 37 species and diverse experimental conditions. CryoDINO learns transferable representations that capture biologically meaningful structural information across molecular and cellular scales. Coupling a shared pretrained backbone with task-specific decoders enables both semantic segmentation and particle detection. Across multiple downstream benchmarks, CryoDINO achieved the strongest overall performance among the evaluated methods and demonstrated high label efficiency, outperforming supervised baselines trained with substantially more annotations. CryoDINO establishes a scalable and transferable representation backbone for generalizable computational pipelines in structural biology. 

\end{abstract}

\section{Introduction}
The development of cryogenic electron tomography (cryo-ET) enabled three-dimensional (3D), near-atomic resolution imaging of biological specimens in their native cellular environments~\cite{ermel2024data,moebel2021deep}. While recent advances enable thousands of cryo-ET tomograms to be generated within days~\cite{peck2024annotating}, their annotation remains time-consuming, laborious, and error-prone~\cite{hecksel2016quantifying}, with considerable reliance on expert manual input~\cite{ermel2025cz}. For example, annotating 29,000 particles across 147 tomograms required 42 experts working for seven days~\cite{peck2024annotating}, underscoring the substantial time and workforce demands associated with tomogram annotation.

\begin{figure}[!t]
  \centering
  \includegraphics[width=0.95\linewidth,height=0.61\textheight,keepaspectratio]{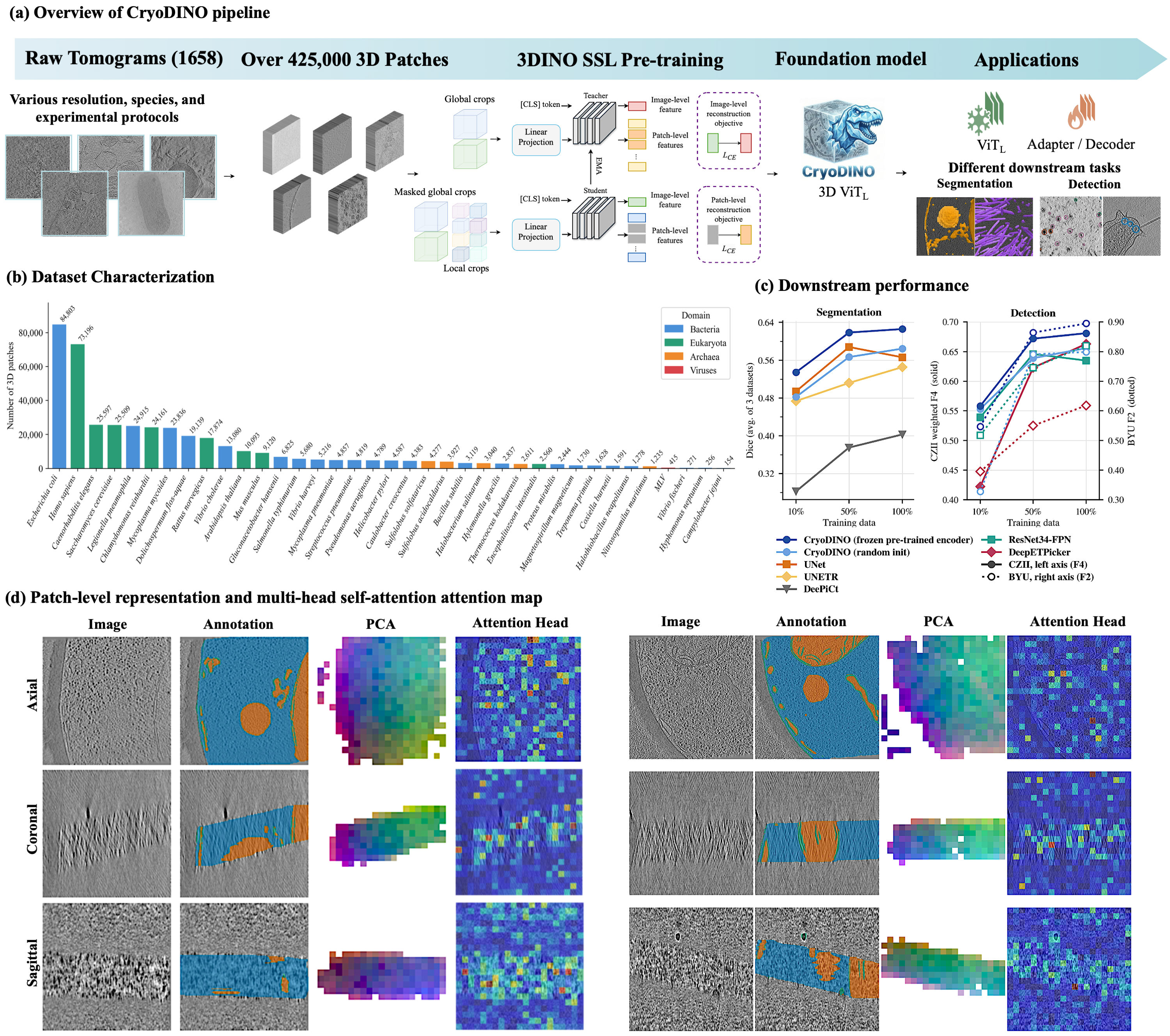}
  \caption{\textbf{Overview of CryoDINO.} (a) A 3D ViT-Large was pretrained with 3DINO~\cite{xu2025generalizable} on 1{,}658 tomograms from 48 datasets and transferred to segmentation and particle detection using task-specific decoders. (b) Distribution of 3D training patches across species, colored by biological domain. (c) Downstream performance with 10\%, 50\%, and 100\% of the training data. CryoDINO uses a frozen pretrained encoder and outperforms the baselines, with the largest gains in the 10\% label regime. (d) Patch-level representations and multi-head self-attention across tomographic planes: raw tomograms, ground truth, PCA feature maps, and final-layer class-token attention.}
  \label{fig:fig1}
\end{figure}

Beyond the challenges of annotation, cryo-ET is constrained by acquisition physics. To reduce radiation damage, the electron dose must be limited, yielding tomograms with low signal-to-noise and contrast-to-noise ratios (SNR and CNR) that obscure structural details and hinder accurate object identification~\cite{liu2024deepetpicker,lamm2026membrain,turk2020promise}. In addition, the restricted tilt range produces the missing wedge artifact, causing anisotropic resolution and orientation-dependent distortions, particularly affecting membranes and macromolecular complexes aligned perpendicular to the electron beam~\cite{liu2024deepetpicker,lamm2026membrain}. Although denoising and missing-wedge correction methods exist~\cite{liu2022isotropic,wiedemann2024deepdewedgetomography}, they may introduce hallucinated structures or bias downstream analysis~\cite{bepler2020topazdenoise,pyle2021current}. As a result, 3D particle localization and structural identification in crowded cellular environments remain more challenging than 2D particle picking in single-particle cryo-EM~\cite{wagner2019crYOLO,bepler2019topaz}, limiting scalable and robust automated cryo-ET analysis~\cite{liu2024deepetpicker,lamm2026membrain,wagner2025cryoet,harastani2025template}.

To facilitate data sharing and accelerate computational analysis, the CryoET Data Portal~\cite{ermel2024data} was established as a large-scale repository of standardized tomograms and metadata. The portal currently hosts more than 60,000 tomograms spanning diverse organisms, cellular environments, and imaging conditions~\cite{cryoetdataportal}. However, despite substantial community-driven annotation efforts, expert molecular annotations remain limited relative to the scale and diversity of the available data~\cite{peck2024annotating,peck2025phantom,peck2026lessons}. This growing abundance of unlabeled tomograms presents an opportunity to learn generalizable representations from cryo-ET data~\cite{ermel2024data,xu2025generalizable}.

Task-specific deep learning approaches have advanced particle picking, macromolecule localization, and segmentation~\cite{de2023convolutional,moebel2021deep,liu2024deepetpicker,ricetomotwin2023,lamm2026membrain}, beyond conventional template matching~\cite{bohm2000template,frangakis2002identification}. However, their reliance on expert annotations limits scalability across datasets, species, and imaging conditions~\cite{ermel2025cz,peck2024annotating,peck2025phantom,peck2026lessons}. Self-supervised learning (SSL) offers a promising alternative by exploiting abundant unlabeled tomograms, yet existing cryo-ET SSL approaches~\cite{stojanovska2025cryosiam,jiang2025fm} have relied on simulated data that cannot fully capture the structural heterogeneity and acquisition variability of real cellular tomograms. Meanwhile, conventional SSL strategies developed for natural images or other 3D domains may be poorly suited to cryo-ET-specific challenges, including low SNR and CNR, and anisotropic sampling. Pretext tasks such as voxel reconstruction or rotation prediction may therefore capture imaging artifacts rather than biologically meaningful structures~\cite{dino,dinov2,xu2025generalizable,munim2026echojepalatentpredictivefoundation,liu2024deepetpicker}. These limitations motivate representation-learning approaches that can learn semantically rich and transferable features directly from real cryo-ET tomograms.

Here, we present CryoDINO, a cryo-ET foundation model pretrained on a large and diverse corpus of 425{,}852 3D patches extracted from 1{,}658 tomograms spanning 37 species, diverse cellular environments, and experimental conditions in the CryoET Data Portal (Fig.~1a--b). Our approach builds on 3DINO~\cite{xu2025generalizable}, a three-dimensional extension of DINOv2~\cite{dino,dinov2} that adapts self-supervised teacher--student distillation to volumetric data. We show that CryoDINO learns transferable, semantically rich representations that capture macromolecular identity and broader cellular context. A shared pretrained backbone, coupled with optimized task-specific decoders, enables both semantic segmentation and particle detection. Across multiple benchmarks, CryoDINO consistently outperforms state-of-the-art task-specific methods while demonstrating strong generalization across species, imaging conditions, and low-annotation regimes (Fig.~1c--d). These results establish CryoDINO as a scalable and label-efficient foundation for general-purpose cryo-ET analysis. Our contributions are threefold:

\begin{itemize}
    \item We introduce CryoDINO, a 3D foundation model for cryo-ET pretrained through self-supervised teacher--student distillation on 425{,}852 unlabeled 3D patches from 1{,}658 tomograms spanning 37 species and diverse imaging conditions and resolutions. The code, pretrained weights, tutorials, visualization tools, and evaluation framework are publicly available at \url{https://github.com/bowang-lab/CryoDINO}.



    \item To the best of our knowledge, CryoDINO is the first cryo-ET foundation model pretrained at scale on \textit{heterogeneous real experimental tomograms}. It learns transferable 3D representations of macromolecular identity and spatial organization that support both semantic segmentation and particle detection through a shared frozen encoder with task-specific decoders.

    \item We demonstrate strong performance across semantic segmentation and particle detection benchmarks, outperforming state-of-the-art task-specific methods while maintaining generalization across acquisition settings and strong label efficiency in low-annotation regimes.

\end{itemize}


\section{Methods}
\subsection{Pretraining and Downstream Datasets}
We curated a large, unlabeled pretraining corpus using data from the CryoET Data Portal~\cite{ermel2024data}, selecting tomograms with high-quality tilt series (score $\geq 4/5$) and excluding tomograms with an axial voxel spacing exceeding $20$\,\AA{} (Fig.~1a--b). To prevent data leakage, acquisitions overlapping with downstream annotated datasets and duplicate runs were removed. To avoid overrepresentation of large studies, each dataset contributed at most 100 training tomograms via random sampling. The resulting corpus spans diverse organisms, cellular contexts, and imaging conditions while limiting dominance by individual large studies. For patch extraction, an approximate foreground mask was generated per tomogram using deconvolution~\cite{tegunov2019warp} followed by Otsu thresholding~\cite{otsu1979threshold}. Mask-guided sampling retained only subvolumes containing more than 4,096 foreground voxels, corresponding to the voxel count of a single 16 × 16 × 16 ViT token, to reduce the prevalence of background-dominated training patches. In total, 425{,}852 biologically meaningful 3D patches were extracted for self-supervised pretraining (Fig.~1a--b).

For downstream evaluation, we considered four segmentation benchmarks with binary or multiclass targets (Table~S1). DS-10001, DS-10010, and EMPIAR-12049 were obtained from the CryoET Data Portal~\cite{ermel2024data}, whereas EMPIAR-10989 was sourced from the Electron Microscopy Public Image Archive (EMPIAR)~\cite{iudin2016empiar}. For CryoET Data Portal datasets, only ground-truth annotations were used, excluding pseudo-labels. Each dataset was split into training, validation, and test sets (Table~S1). To evaluate performance under limited supervision, we additionally trained using randomly sampled 50\% and 10\% subsets of the training data while keeping validation and test sets fixed. For particle detection, we benchmark on the CZII CryoET Object Identification Challenge phantom dataset~\cite{peck2025phantom,peck2026lessons} (DS-10440 for training, DS-10445 for validation, and DS-10446 for testing) and the BYU bacterial flagellar motor dataset from the Kaggle challenge~\cite{owens2025motorbench}, excluding the tomograms without positive motor annotations. 
Datasets and split details are provided in Table~S2. 
\subsection{Pretraining Methods}
Our implementation builds on 3DINO~\cite{xu2025generalizable}, a volumetric extension of DINOv2~\cite{dino,dinov2}, using the official DINOv2 PyTorch codebase. Data loading and augmentation were implemented with MONAI~\cite{cardoso2022monai}. Pretraining was performed on four NVIDIA B200-183GB GPUs with a ViT-Large backbone and a patch size of $16 \times 16 \times 16$. Standard pretraining used a batch size of 275 per GPU (1{,}100 total), global crops of $96^3$, and eight local crops of $48^3$. The base learning rate was 0.002, and the teacher momentum $\tau$ increased from 0.992 to 1.000 via cosine scheduling. Training ran for 125{,}000 iterations (\textasciitilde{5} days). A subsequent high-resolution adaptation stage followed the 3DINO protocol with identical scheduling compressed to 12{,}500 iterations. High-resolution adaptation used global crops of $128^3$, local crops of $64^3$, a batch size of 130 per GPU (520 total), and a base learning rate of 0.001. This stage was initialized from the stage-1 checkpoint at iteration 112{,}500 and required approximately 16 hours. Pre-extracted patches were loaded at native resolution and independently normalized with one of two randomly selected schemes, applied with equal probability to simulate acquisition variability: percentile-based intensity clipping, with bounds randomly sampled from predefined percentile pairs ($2$--$98$, $1$--$99$, $0.5$--$99.5$, $0.05$--$99.95$ and $0.02$--$99.98$), or tomogram-level $z$-scoring in which the patch is centered on the tomogram mean and divided by $k$ times the tomogram standard deviation, with $k$ drawn uniformly from $\{1, 2, 3, 4, 5\}$, followed by clipping to $[-1, 1]$. Geometric augmentations included 3D random resized cropping (with in-plane scaling), random flips, and $90^\circ$ rotations along all axes. Intensity augmentations comprised random contrast adjustment, Gaussian smoothing and sharpening, intensity scaling and shifting, and Gibbs and Gaussian noise~\cite{xu2025generalizable,cardoso2022monai}.

\subsection{Finetuning Methods}
We first performed linear probing~\cite{dino,dinov2} to evaluate representation quality and select the optimal pretrained checkpoint. A lightweight two-layer channel-wise linear head was attached to the frozen ViT backbone. We evaluated every saved checkpoint and a randomly initialized backbone on the segmentation datasets. The linear head was trained for 100 epochs (12{,}500 iterations) using AdamW~\cite{loshchilov2017decoupled} (learning rate 0.001, cosine schedule, 3{,}000 warm-up iterations, batch size 2). The checkpoint achieving the highest validation Dice was selected for subsequent experiments. 
Using the selected backbone, we evaluated ViTAdapterUNETR~\cite{chen2022vision,xu2025generalizable}, under two regimes: (i) frozen backbone (only the adapter and decoder are trained) and (ii) full fine-tuning (joint encoder–decoder optimization). Fine-tuning employed a batch size of 2, learning rate $1\times10^{-4}$ with cosine decay and 3{,}000 warm-up iterations, and DiceCE loss for 100 epochs (30{,}000 iterations). To assess label efficiency, we additionally trained with 10\% and 50\% of labeled data while keeping all hyperparameters fixed. Training used pre-extracted $512^3$ patches at native resolution with tomogram-level z-score normalization. Four random $128^3$ sub-volumes were sampled per patch (3:1 foreground-to-background ratio), followed by percentile-based normalization to $[-1,1]$ and the same augmentations as in pretraining. Inference used sliding-window evaluation ($128^3$ window, 0.75 overlap).


For segmentation, we compared CryoDINO with an identical randomly initialized model, 3D U-Net~\cite{ronneberger2015u} and UNETR~\cite{hatamizadeh2022unetr} implemented in MONAI~\cite{cardoso2022monai}, DeePiCt~\cite{de2023convolutional}, and MemBrain-seg~\cite{lamm2026membrain}. All models were trained for 30{,}000 iterations using identical data splits and evaluated with foreground Dice and the 95th-percentile Hausdorff distance (HD95)~\cite{taha2015metrics}.


For particle picking, CryoDINO was paired with the ViTAdapterUNETR decoder with feature pyramid network (FPN) fusion~\cite{lin2017fpn}. The four multi-scale feature maps from the adapter were fused top-down through three UNETR up-sampling blocks, stopping at the penultimate level, half the input resolution~\cite{daddies2025czii}. The anchor-free detection head predicts, at each output cell $i$ with position $\mathbf{a}_i$, per-class center scores $\hat q_{i,c}$ and a 3D offset $\mathbf{o}_i$, giving a predicted center $\hat{\mathbf{p}}_i=\mathbf{a}_i+\mathbf{o}_i$. It was trained with a varifocal classification loss~\cite{zhang2021varifocal,daddies2025czii} and an IoU-based center-regression loss~\cite{daddies2025czii}. Because particles are points, the IoU is replaced by a Gaussian keypoint similarity $S(\mathbf{p},\mathbf{g})=\exp\!\left(-\lVert\mathbf{p}-\mathbf{g}\rVert^2/2\sigma_g^2\right)$ between a predicted center $\mathbf{p}$ and a ground-truth particle center $\mathbf{g}$, where $\sigma_g$ is set by the size of particle $\mathbf{g}$. A task-aligned assigner~\cite{feng2021tood} selects, for each particle, up to 13 nearby cells with the highest alignment score $\hat q_{i,c_g}\,S(\hat{\mathbf{p}}_i,\mathbf{g})^6$ as positives and gives each a soft target $t_i\in[0,1]$ proportional to that score. The classification loss uses binary cross-entropy, weighting positives by $t_i$ and negatives by $0.75\,\hat q_{i,c}^2$ to suppress easy background cells. The regression loss is $\sum_{i\in\mathcal{P}} t_i\left(1-S(\hat{\mathbf{p}}_i,\mathbf{g}_i)\right)$. Both losses are normalized by $\sum_i t_i$ and summed. We fine-tuned the model in two settings: (i) with the encoder frozen, to assess the quality of CryoDINO's learned representations, and (ii) with low-rank adaptation (LoRA with rank 16), to assess its adaptability to downstream particle-picking tasks. At inference, stride-two outputs were assigned their highest-scoring class, and predicted offsets defined candidate centers. Class-wise non-maximum suppression (NMS) removed lower-scoring candidates whose overlap with a retained candidate exceeded 0.8~\cite{daddies2025czii}. We selected class-specific score thresholds by maximizing validation metrics, which are weighted F4 on CZII dataset and F2 on BYU dataset, and applied them unchanged to the test tomograms. 


Checkpoints were selected on the official validation set and evaluated on the test set (Table~S2). For CZII dataset, performance was measured using the class-weighted F4 score, with predictions within half the particle radius counted as true positives~\cite{peck2024annotating}. Class weights were 2 for $\beta$-galactosidase and thyroglobulin, 1 for apo-ferritin, ribosome, and virus-like particle, and 0 for $\beta$-amylase. Low-data experiments used four and one tomograms from the training set (50\% and 10\%, respectively). For the BYU dataset, we used the F2 metric with a matching distance of 1,000~\AA{}. Both F4 and F2 were computed based on the official metric implementation from the CZII CryoET Object Identification Challenge~\cite{peck2025phantom,peck2026lessons}. We evaluated CryoDINO both with pretrained weights and from random initialization, alongside DeepETPicker trained at the tomogram level~\cite{liu2024deepetpicker} and a RetinaNet with a ResNet34-FPN backbone trained from scratch (MONAI implementation~\cite{cardoso2022monai,monai_detection_tutorial}, RetinaNet~\cite{lin2017focal}). Each model was trained for 30{,}000 iterations and evaluated with the same detection metrics. We report 95\% bootstrap confidence intervals over test tomograms~\cite{efron1979bootstrap}.

\section{Results}

\subsection{Linear Probing}

We evaluated the patch-level representations of CryoDINO with a frozen-backbone linear probing protocol and compared them against a randomly initialized ViT-Large~\cite{dosovitskiy2020image}. Fig.~S1 reports the foreground validation Dice on DS-10001, DS-10010 and EMPIAR-12049, and their average, for every evaluated checkpoint. EMPIAR-10989, which has a single training tomogram, is omitted because a linear head collapsed to background for every backbone, including the randomly initialized one. Probing the pretrained weights outperformed random initialization at every checkpoint and improved during stage-1 pretraining (96$^3$ crops): from 0.68 at 12.5K iterations to 0.78 at 75K iterations on DS-10001 (random initialization: 0.32), and from 0.02 to 0.10 on EMPIAR-12049 (random initialization: 0.01). Performance plateaued on DS-10001, kept rising slowly on EMPIAR-12049 and fluctuated between 0.18 and 0.32 on DS-10010 (random initialization: 0.07). The high-resolution adaptation stage further improved linear-probe performance, with the selected checkpoint achieving the highest mean validation Dice across datasets. We selected the 6{,}249-iteration checkpoint of the high-resolution stage for all fine-tuning experiments, based on the highest validation Dice across datasets; with a linear head it reached 0.74 on DS-10001 (vs.\ 0.32 for random initialization), 0.30 on DS-10010 (vs.\ 0.07) and 0.14 on EMPIAR-12049 (vs.\ 0.01) (Fig.~S1).

To inspect the learned embeddings we visualized patch-level principal components (PCA) and multi-head self-attention (MHSA) maps of the frozen backbone (Fig.~1d and Fig.~S2). The PCA projections separated the foreground cellular context from the background on unseen downstream data and grouped structurally similar regions. For DS-10001, components inside the foreground (cytoplasm, organelles, membrane) were distinct from the background. MHSA maps concentrated on structured regions despite acquisition noise and reconstruction artifacts.

\subsection{CryoDINO enables generalizable and label-efficient semantic segmentation}

We compared CryoDINO with four supervised baselines trained from scratch on the same datasets: (i) a five-level residual 3D U-Net and (ii) UNETR with a ViT-Base encoder~\cite{hatamizadeh2022unetr}, both implemented in MONAI~\cite{cardoso2022monai}; (iii) DeePiCt~\cite{de2023convolutional}, trained using its default cryo-ET segmentation pipeline; and (iv) MemBrain-seg~\cite{lamm2026membrain}, a 3D U-Net model for membrane classes. To isolate the effect of pretraining, we additionally trained the same ViTAdapterUNETR architecture with a randomly initialized, fully trainable encoder. All methods used identical training, validation, and test splits. Dice and HD95 were computed for each held-out tomogram and averaged across foreground classes.

With all labels (100\%), CryoDINO obtained the highest foreground Dice on all four datasets (Table~1 and Fig.~S3--6): 0.858 on DS-10001, 0.697 on DS-10010, 0.354 on EMPIAR-10989 and 0.322 on EMPIAR-12049, compared with 0.807, 0.641, 0.334, and 0.321, respectively, for the strongest competing method on each dataset. No single baseline was consistently second best: the runner-up alternated between U-Net, UNETR and the randomly initialized model across datasets (Table~1 and Fig.~S3--4). CryoDINO also had the lowest HD95 on DS-10001 (23.5 vs.\ 47.0 for the next best), DS-10010 (213.9 vs.\ 260.7) and EMPIAR-12049 (115.7 vs.\ 139.0). On EMPIAR-10989 the random-initialization model was marginally lower (22.0 vs.\ 23.1). Comparison with an architecturally identical model trained from scratch highlights the benefit of pretrained CryoDINO representations over fully supervised learning without pretraining. CryoDINO produced substantial improvements on DS-10001 (+0.051 Dice; HD95 reduced from 67.0 to 23.5) and DS-10010 (+0.071 Dice), a smaller improvement on EMPIAR-10989 (+0.020), and a negligible improvement on EMPIAR-12049 (+0.001). EMPIAR-12049 is the most class-imbalanced dataset, and all methods achieved foreground Dice scores below 0.33. DeePiCt~\cite{de2023convolutional}, trained using its default multiclass configuration, achieved 0.473 Dice on DS-10001, largely because of poor organelle segmentation (0.053 Dice), and was competitive only on EMPIAR-12049 (0.258). On membrane classes alone, CryoDINO outperformed MemBrain-seg~\cite{lamm2026membrain} on all three applicable datasets, achieving Dice scores of 0.721 vs.\ 0.638, 0.697 vs.\ 0.524, and 0.333 vs.\ 0.265 (Fig.~2b, Table~1b, and Fig.~S6). Qualitatively, CryoDINO more accurately segmented organelles, produced fewer false-positive ribosome predictions, and better preserved the continuity of membranes and microtubules, whereas competing methods struggled to recover these structures (Fig.~2 and Figs.~S5--S6).

\providecommand{\sd}[1]{{\scriptsize\,$\pm$\,#1}}
\providecommand{\dsname}[2]{\raisebox{-#1\normalbaselineskip}[0pt][0pt]{#2}}

\begin{table}[!tb]
\centering
\footnotesize
\setlength{\tabcolsep}{5pt}
\caption{\textbf{Segmentation results on the test sets.} Dice and HD95 (mean\,$\pm$\,s.d.\ over test tomograms) when training on 10\%, 50\% and 100\% of the training tomograms. Best value per column and dataset in bold, second best underlined. (a)~Foreground classes averaged, background excluded. (b)~Membrane class only, including the comparison with MemBrain-seg. EMPIAR-10989 has a single training tomogram, so only the 100\% setting is reported.}
{\raggedright\textbf{(a)} Foreground classes (background excluded), averaged over the classes of each dataset\par}
\smallskip
\resizebox{\linewidth}{!}{%
\begin{tabular}{@{}ll ccc ccc@{}}
\toprule
 & & \multicolumn{3}{c}{Foreground Dice $\uparrow$} & \multicolumn{3}{c}{Foreground HD95 $\downarrow$} \\
\cmidrule(lr){3-5} \cmidrule(l){6-8}
Dataset & Method & 10\% & 50\% & 100\% & 10\% & 50\% & 100\% \\
\midrule
\dsname{2}{DS-10001} & U-Net & \underline{0.786}\sd{0.090} & \underline{0.828}\sd{0.097} & 0.799\sd{0.119} & 52.1\sd{45.5} & \underline{58.1}\sd{68.3} & \underline{47.0}\sd{50.4} \\
 & UNETR & 0.700\sd{0.128} & 0.769\sd{0.119} & 0.773\sd{0.128} & 59.7\sd{47.1} & \underline{58.1}\sd{54.8} & 51.1\sd{46.7} \\
 & DeePiCt & 0.401\sd{0.031} & 0.436\sd{0.008} & 0.473\sd{0.004} & 241.8\sd{110.1} & 294.6\sd{165.5} & 298.4\sd{172.1} \\
 & CryoDINO (random init.) & 0.719\sd{0.151} & 0.774\sd{0.125} & \underline{0.807}\sd{0.098} & \underline{51.2}\sd{35.5} & 67.2\sd{62.4} & 67.0\sd{64.8} \\
 & CryoDINO (frozen encoder) & \textbf{0.820}\sd{0.096} & \textbf{0.855}\sd{0.086} & \textbf{0.858}\sd{0.077} & \textbf{47.8}\sd{39.8} & \textbf{24.0}\sd{23.3} & \textbf{23.5}\sd{22.3} \\
\midrule
\dsname{2}{EMPIAR-10989} & U-Net & -- & -- & 0.256 & -- & -- & 23.3 \\
 & UNETR & -- & -- & 0.301 & -- & -- & 31.5 \\
 & DeePiCt & -- & -- & 0.127 & -- & -- & 39.1 \\
 & CryoDINO (random init.) & -- & -- & \underline{0.334} & -- & -- & \textbf{22.0} \\
 & CryoDINO (frozen encoder) & -- & -- & \textbf{0.354} & -- & -- & \underline{23.1} \\
\midrule
\dsname{2}{EMPIAR-12049} & U-Net & 0.067\sd{0.032} & 0.293\sd{0.341} & 0.303\sd{0.348} & \underline{123.4}\sd{38.6} & \underline{148.2}\sd{114.6} & \underline{139.0}\sd{117.6} \\
 & UNETR & 0.092\sd{0.067} & 0.153\sd{0.147} & 0.222\sd{0.238} & 128.4\sd{56.8} & 215.8\sd{153.5} & 195.6\sd{174.1} \\
 & DeePiCt & 0.014\sd{0.012} & 0.166\sd{0.166} & 0.258\sd{0.290} & 262.3\sd{46.7} & 227.9\sd{141.6} & 198.4\sd{112.1} \\
 & CryoDINO (random init.) & \underline{0.113}\sd{0.095} & \underline{0.312}\sd{0.364} & \underline{0.321}\sd{0.376} & 138.2\sd{48.0} & 188.2\sd{144.8} & 176.7\sd{150.7} \\
 & CryoDINO (frozen encoder) & \textbf{0.128}\sd{0.118} & \textbf{0.326}\sd{0.376} & \textbf{0.322}\sd{0.379} & \textbf{109.6}\sd{67.9} & \textbf{130.8}\sd{112.4} & \textbf{115.7}\sd{108.6} \\
\bottomrule
\end{tabular}}

\vspace{4pt}
{\raggedright\textbf{(b)} Membrane class only (all methods scored on the same class, including the membrane-only model MemBrain-seg)\par}
\smallskip
\resizebox{\linewidth}{!}{%
\begin{tabular}{@{}ll ccc ccc@{}}
\toprule
 & & \multicolumn{3}{c}{Membrane Dice $\uparrow$} & \multicolumn{3}{c}{Membrane HD95 $\downarrow$} \\
\cmidrule(lr){3-5} \cmidrule(l){6-8}
Dataset & Method & 10\% & 50\% & 100\% & 10\% & 50\% & 100\% \\
\midrule
\dsname{2.5}{DS-10001} & U-Net & \underline{0.672}\sd{0.177} & \underline{0.692}\sd{0.173} & 0.673\sd{0.168} & 44.8\sd{56.8} & \underline{46.0}\sd{57.0} & \textbf{42.7}\sd{52.3} \\
 & UNETR & 0.569\sd{0.252} & 0.637\sd{0.237} & 0.663\sd{0.197} & 48.7\sd{54.2} & 47.6\sd{56.2} & \underline{43.0}\sd{51.3} \\
 & DeePiCt & 0.657\sd{0.260} & 0.591\sd{0.261} & 0.684\sd{0.146} & 72.2\sd{2.8} & 197.3\sd{175.7} & 207.4\sd{186.6} \\
 & MemBrain-seg & 0.484\sd{0.099} & 0.609\sd{0.104} & 0.638\sd{0.137} & 53.8\sd{46.4} & 96.1\sd{6.7} & 62.3\sd{31.2} \\
 & CryoDINO (random init.) & 0.669\sd{0.189} & 0.682\sd{0.162} & \underline{0.694}\sd{0.160} & \underline{44.7}\sd{54.3} & 46.1\sd{55.2} & 45.4\sd{53.4} \\
 & CryoDINO (frozen encoder) & \textbf{0.712}\sd{0.170} & \textbf{0.719}\sd{0.168} & \textbf{0.721}\sd{0.164} & \textbf{43.3}\sd{52.3} & \textbf{43.5}\sd{53.6} & 43.5\sd{54.6} \\
\midrule
\dsname{2.5}{DS-10010} & U-Net & \underline{0.628}\sd{0.112} & \underline{0.643}\sd{0.121} & 0.597\sd{0.116} & 271.2\sd{177.9} & 254.9\sd{188.5} & 265.2\sd{166.3} \\
 & UNETR & \underline{0.628}\sd{0.136} & 0.614\sd{0.135} & \underline{0.641}\sd{0.116} & \underline{246.7}\sd{198.0} & 259.3\sd{189.8} & \underline{260.7}\sd{185.4} \\
 & DeePiCt & 0.432\sd{0.220} & 0.523\sd{0.158} & 0.478\sd{0.236} & 277.3\sd{174.7} & 274.3\sd{183.0} & 276.9\sd{178.3} \\
 & MemBrain-seg & 0.519\sd{0.158} & 0.529\sd{0.153} & 0.524\sd{0.152} & 285.7\sd{169.1} & 284.9\sd{169.4} & 284.6\sd{169.6} \\
 & CryoDINO (random init.) & 0.615\sd{0.135} & 0.614\sd{0.135} & 0.626\sd{0.137} & 253.5\sd{197.8} & \underline{244.1}\sd{205.1} & 268.7\sd{182.6} \\
 & CryoDINO (frozen encoder) & \textbf{0.655}\sd{0.093} & \textbf{0.675}\sd{0.103} & \textbf{0.697}\sd{0.073} & \textbf{224.6}\sd{227.8} & \textbf{228.5}\sd{218.6} & \textbf{213.9}\sd{212.6} \\
\midrule
\dsname{2.5}{EMPIAR-12049} & U-Net & 0.083\sd{0.019} & 0.327\sd{0.335} & 0.318\sd{0.320} & \underline{172.3}\sd{76.8} & 177.0\sd{108.4} & 168.6\sd{126.0} \\
 & UNETR & 0.121\sd{0.035} & 0.234\sd{0.200} & 0.255\sd{0.233} & 180.2\sd{70.0} & \underline{172.3}\sd{99.7} & 168.1\sd{89.8} \\
 & DeePiCt & 0.014\sd{0.018} & 0.168\sd{0.128} & 0.209\sd{0.175} & 220.7\sd{67.9} & 214.0\sd{141.4} & 235.9\sd{72.6} \\
 & MemBrain-seg & \textbf{0.214}\sd{0.167} & 0.243\sd{0.244} & 0.265\sd{0.258} & 228.6\sd{55.8} & 302.6\sd{112.5} & 274.6\sd{96.4} \\
 & CryoDINO (random init.) & 0.130\sd{0.041} & \underline{0.341}\sd{0.343} & \textbf{0.345}\sd{0.359} & 191.2\sd{49.9} & 173.8\sd{122.4} & \underline{159.8}\sd{117.5} \\
 & CryoDINO (frozen encoder) & \underline{0.155}\sd{0.080} & \textbf{0.349}\sd{0.351} & \underline{0.333}\sd{0.344} & \textbf{138.6}\sd{48.2} & \textbf{127.3}\sd{111.5} & \textbf{119.5}\sd{126.7} \\
\bottomrule
\end{tabular}}
\label{tab:seg_results}
\end{table}

\begin{figure}[!t]
  \centering
  \includegraphics[width=0.90\textwidth,height=0.56\textheight,keepaspectratio]{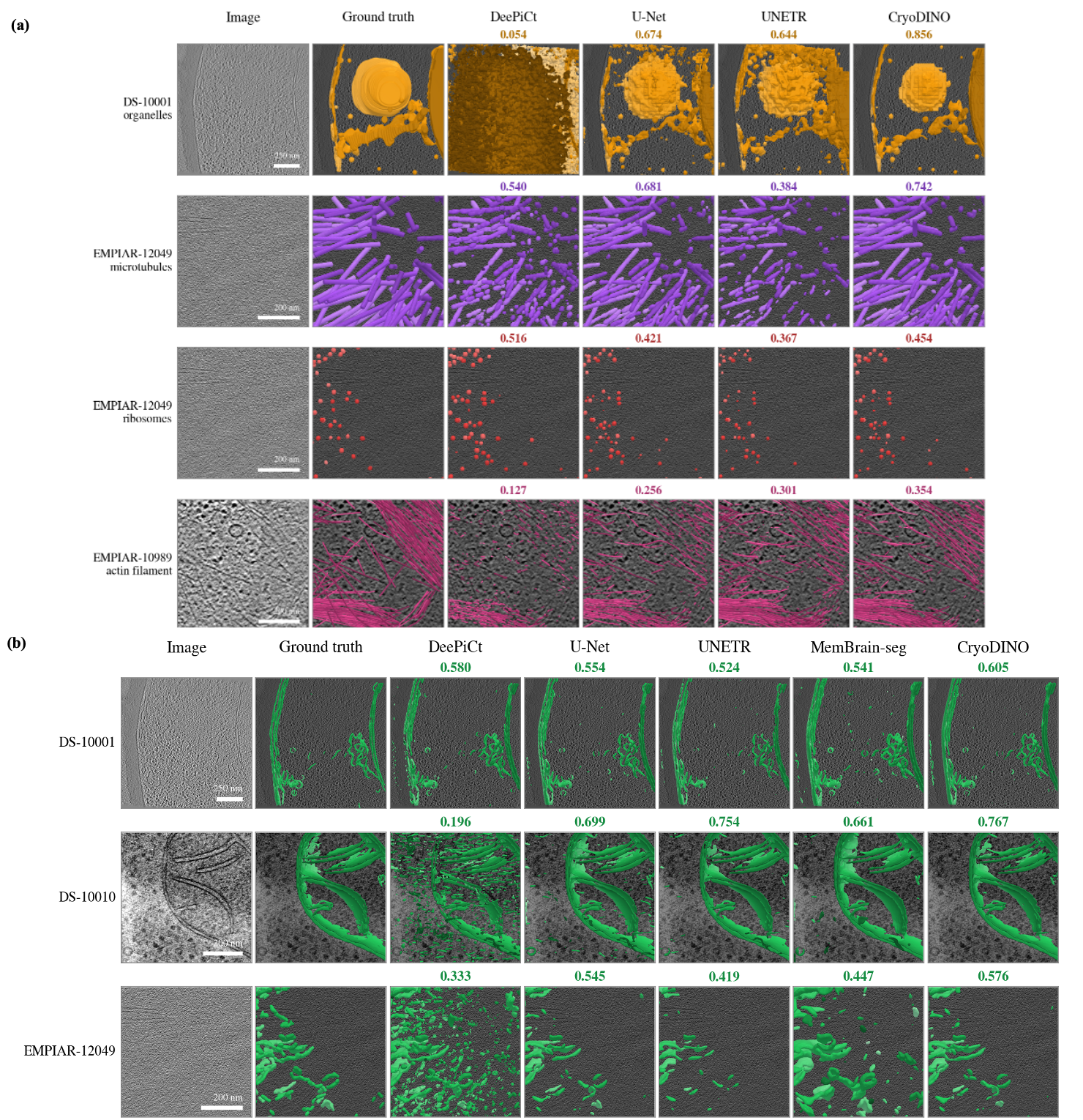}
  \caption{\textbf{Qualitative comparison of segmentation performance.} 3D representative multiclass (a) and membrane (b) predictions overlaid on tomographic slices, with class-wise Dice scores shown above each prediction. CryoDINO (frozen encoder) better preserves the morphology and spatial organization of the ground-truth structures, achieving the highest Dice in six of the seven examples.}
  \label{fig:labeleff}
\end{figure}

CryoDINO retained strong segmentation performance when trained with limited annotations (Table~1, Fig.~1c, and Fig.~S3). With a single training tomogram (10\%) on DS-10001 it reached 0.820 Dice, above U-Net~\cite{cardoso2022monai}, UNETR~\cite{hatamizadeh2022unetr}, and the randomly initialized model trained on all eight (0.799, 0.773, 0.807). On DS-10010, 10\% of the labels (three tomograms) gave 0.655, higher than every baseline trained on 100\% (best 0.641). On EMPIAR-12049, half of the labels (four tomograms) gave 0.326, on par with the best baseline trained on all of them (0.321), while a single tomogram yielded low performance across all multiclass methods (Dice 0.014--0.128). The pretraining gain on DS-10001 was largest in the low-label regime (+0.101 Dice at 10\% vs.\ +0.051 at 100\%). Averaged across the three datasets evaluated at 10\%, 50\%, and 100\% label fractions (Fig.~1c), CryoDINO achieved the highest foreground Dice at every fraction (0.534, 0.619, and 0.626, respectively). Averaged across the three datasets, CryoDINO with only half of the labels achieved a higher foreground Dice than every baseline trained on all available labels (0.619 vs.\ 0.585 for the strongest baseline), with only a modest additional gain at 100\% (+0.007). CryoDINO also achieved the lowest HD95 at every label fraction across all three datasets (Fig.~S3).

Because each dataset contained only one to five held-out tomograms, we pooled the ten test tomograms across datasets and applied paired Wilcoxon signed-rank tests~\cite{wilcoxon1945}. At the 100\% label setting, CryoDINO achieved higher Dice and lower HD95 than DeePiCt~\cite{de2023convolutional} and UNETR~\cite{hatamizadeh2022unetr} on all ten tomograms ($p=0.002$; Holm-adjusted $p=0.047$ across 24 comparisons)~\cite{holm1979}. CryoDINO outperformed U-Net~\cite{cardoso2022monai} and random initialization on nine of ten tomograms (Dice $p=0.004$; HD95 $p=0.004$--$0.014$), although these comparisons did not remain significant after multiple-comparison correction (adjusted $p=0.078$--$0.137$). Additional held-out tomograms are required to establish statistical significance within individual datasets.

We next examined whether end-to-end fine-tuning could further improve segmentation performance. Fully fine-tuning the pretrained encoder did not improve over keeping it frozen on any of the four datasets: 0.840 vs.\ 0.858 on DS-10001, 0.671 vs.\ 0.697 on DS-10010, 0.306 vs.\ 0.354 on EMPIAR-10989 and 0.295 vs.\ 0.322 on EMPIAR-12049. This may be explained by the limited number of labeled training samples, as updating all 350 million encoder parameters may increase susceptibility to overfitting. These results indicate that CryoDINO representations are already well aligned with cryo-ET structure and require little task-specific adaptation for segmentation.

\subsection{CryoDINO transfers effectively across particle-detection benchmarks}
\begin{figure}[!htbp]
  \centering
  \includegraphics[width=0.95\textwidth,height=0.80\textheight,keepaspectratio]{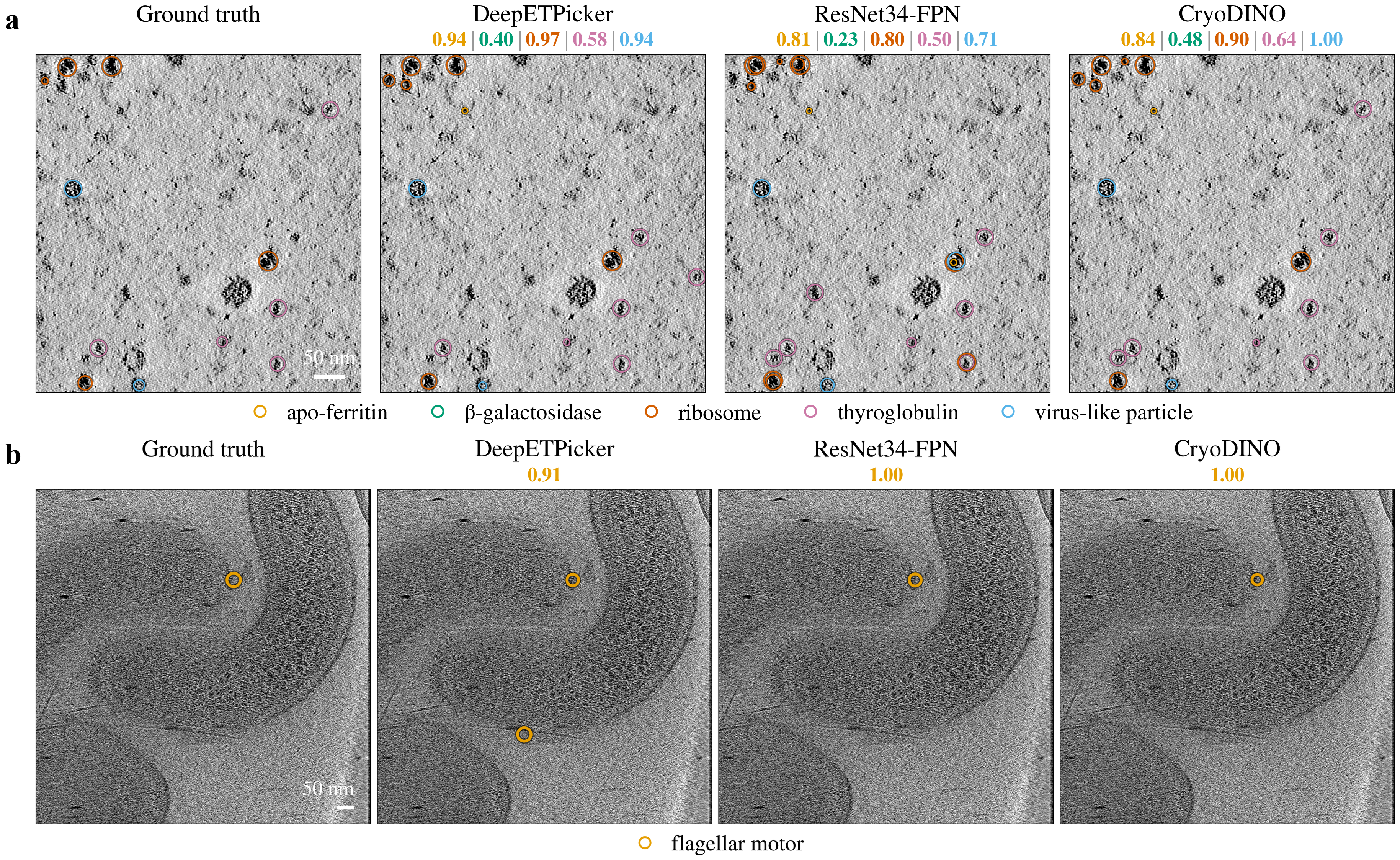}
  \caption{\textbf{Qualitative particle-detection results.} Representative test tomograms from (a) CZII and (b) BYU datasets. Both panels show ground-truth locations and predictions from DeepETPicker, ResNet34-FPN, and CryoDINO (frozen encoder). Numbers indicate class-specific F4 scores for the displayed CZII tomogram (a) and tomogram-level F2 scores for BYU (b).}
  \label{fig:particle_detection}
\end{figure}

To evaluate transfer beyond semantic segmentation, we coupled the frozen CryoDINO encoder with an optimized decoder for particle detection. We compared CryoDINO with DeepETPicker~\cite{liu2024deepetpicker}, ResNet34-FPN, and the architecturally identical detection model trained from random initialization. Across experiments using 10\%, 50\%, and 100\% of the training data, CryoDINO achieved the highest aggregate detection performance at every label fraction on both benchmarks (Table~\ref{tab:czii_detection}, Fig.~1c, and Fig.~S7-8) and produced more precise particle localization (Fig.~\ref{fig:particle_detection} and Fig.~S9).

\begin{table*}[t]
\centering
\footnotesize
\setlength{\tabcolsep}{3pt}
\smallskip
\caption{\textbf{Particle-detection performance on the CZII and BYU datasets.} CZII: weighted F4 with 95\% bootstrap confidence intervals over the 364 test tomograms (training on 1, 4 and 7 tomograms). BYU flagellar motor: F2 (1000~\AA{} matching distance) with 95\% bootstrap confidence intervals over the 55 test tomograms (training on 26, 127 and 253 tomograms). Columns give the fraction of training tomograms (10\%, 50\%, 100\%); thresholds selected on the validation set.}
\label{tab:czii_detection}
\resizebox{\textwidth}{!}{%
\begin{tabular}{@{}ll ccc@{}}
\toprule
 & & \multicolumn{3}{c}{F4 (CZII) / F2 (BYU) $\uparrow$} \\
\cmidrule(l){3-5}
Dataset & Method & 10\% & 50\% & 100\% \\
\midrule
CZII & DeepETPicker & 0.422\ci{0.410}{0.432} & 0.623\ci{0.613}{0.633} & \underline{0.663}\ci{0.651}{0.674} \\
 & ResNet34-FPN & 0.539\ci{0.528}{0.550} & \underline{0.646}\ci{0.635}{0.656} & 0.635\ci{0.623}{0.646} \\
 & CryoDINO (random init.) & \underline{0.553}\ci{0.540}{0.564} & 0.639\ci{0.628}{0.650} & 0.656\ci{0.645}{0.666} \\
 & CryoDINO (frozen encoder) & \textbf{0.558}\ci{0.545}{0.570} & \textbf{0.672}\ci{0.661}{0.682} & \textbf{0.681}\ci{0.670}{0.691} \\
\midrule
BYU & DeepETPicker & 0.395\ci{0.294}{0.502} & 0.550\ci{0.386}{0.723} & 0.618\ci{0.517}{0.716} \\
 & ResNet34-FPN & \underline{0.518}\ci{0.382}{0.646} & 0.746\ci{0.638}{0.844} & \underline{0.820}\ci{0.708}{0.919} \\
 & CryoDINO (random init.) & 0.328\ci{0.204}{0.452} & \underline{0.791}\ci{0.683}{0.889} & 0.799\ci{0.688}{0.900} \\
 & CryoDINO (frozen encoder) & \textbf{0.547}\ci{0.439}{0.655} & \textbf{0.864}\ci{0.777}{0.939} & \textbf{0.895}\ci{0.804}{0.969} \\
\bottomrule
\end{tabular}%
}
\end{table*}

On CZII, the F4 score increased from 0.558 at 10\% to 0.681 at 100\% training data, while on BYU the F2 score increased from 0.547 to 0.895. Notably, the advantage over DeepETPicker was largest in the low-annotation setting on both benchmarks, reaching +0.136 F4 on CZII and +0.314 F2 on BYU at 10\% and 50\% of the training data, respectively. Self-supervised pretraining consistently improved performance. The largest gain over random initialization was observed at 10\% of the training data on BYU (+0.219 F2), whereas on CZII the largest gain occurred at 50\% of the training data (+0.033 F4), highlighting the benefits of both pretraining and label efficiency.

On CZII, the largest class-level gains were observed for $\beta$-galactosidase and thyroglobulin, the two challenging particle classes assigned double weight in the aggregate F4 score. Averaged across the three training-set sizes, CryoDINO achieved an F4 score of 0.471 for $\beta$-galactosidase, compared with 0.440 for random initialization, 0.424 for ResNet34-FPN, and 0.394 for DeepETPicker~\cite{liu2024deepetpicker}, and ranked first at both the 50\% and 100\% settings. For thyroglobulin, CryoDINO achieved 0.542, exceeding DeepETPicker (0.529), random initialization (0.512), and ResNet34-FPN (0.506). These results suggest that the largest gains from pretraining arise for the most challenging particle classes. 


We next examined whether adapting the pretrained encoder could further improve detection performance. End-to-end fine-tuning did not provide additional gains, whereas LoRA~\cite{hu2022lora} increased the CZII F4 score from 0.681 to 0.690 and the BYU F2 score from 0.895 to 0.915. These modest improvements indicate that frozen CryoDINO representations transfer effectively across particle-detection benchmarks while retaining the capacity for parameter-efficient task-specific adaptation.

\section{Discussion and Conclusion}

We introduce CryoDINO, a general-purpose 3D foundation model pretrained on a large and diverse corpus of unlabeled cryo-ET tomograms through teacher--student self-distillation. Across multiple semantic segmentation and particle detection benchmarks spanning diverse species, imaging conditions, and resolutions, CryoDINO consistently achieved the strongest overall performance among the evaluated methods while demonstrating high label efficiency. Linear probing, attention-map visualization, and comparisons with an architecturally identical model trained from random initialization show that CryoDINO learns semantically meaningful 3D representations that capture structural and contextual information relevant to macromolecular identification and spatial organization during pretraining. Moreover, coupling the frozen encoder with task-specific segmentation and detection decoders demonstrates that these representations transfer effectively across distinct downstream tasks while substantially reducing reliance on expert annotations.

When evaluated across multiple downstream tasks, CryoDINO exhibited the strongest benefits in low-annotation regimes, consistent with observations from other large-scale self-supervised learning studies~\cite{dinov2,xu2025generalizable,munim2026echojepalatentpredictivefoundation}. Across both semantic segmentation and particle detection benchmarks, CryoDINO frequently matched or exceeded task-specific supervised methods trained with all available annotations while using only 10\% or 50\% of the labeled tomograms. In some cases, 10\% corresponded to a single training tomogram. The largest gains were observed on the most challenging datasets and particle classes, including $\beta$-galactosidase and thyroglobulin in CZII and flagellar-motor detection in BYU. Comparisons with architecturally identical models trained from random initialization further demonstrated that these improvements arose from the learned representations rather than the model architecture alone, highlighting the value of large-scale self-supervised pretraining on diverse cryo-ET data.

While CryoDINO demonstrates strong performance and generalization, further optimization of hyperparameters and training strategies for cryo-ET-specific properties, including anisotropic resolution~\cite{liu2022isotropic}, low signal-to-noise ratios, and severe class imbalance, may yield additional gains. The pretrained backbone used throughout this study was selected using segmentation validation performance alone. Although a different pretraining checkpoint may prove optimal for individual detection tasks, this checkpoint transferred effectively across both segmentation and particle detection, suggesting that the learned representations generalize across downstream tasks. For CZII particle detection, model checkpoints were selected during fine-tuning using the challenge metric~\cite{peck2025phantom,peck2026lessons}, because available ground-truth annotations may underrepresent the true number of particles~\cite{peck2025phantom,peck2026lessons}. Alternative loss functions and model-selection metrics may be more appropriate for applications in which precision and recall are equally important. Extending CryoDINO to additional particle-picking settings~\cite{bendory2020single}, subtomogram averaging, and 3D macromolecular reconstruction would further characterize its utility across the cryo-ET analysis pipeline. Finally, integrating nnU-Net--style post-processing strategies~\cite{isensee2021nnu} may further improve performance.

Overall, our results establish CryoDINO as a general-purpose foundation model for cryo-ET. By decoupling large-scale representation learning from dense supervision and supporting both semantic segmentation and particle detection through task-specific decoders, CryoDINO provides a scalable and transferable foundation for automated cryo-ET analysis. To facilitate future research, we release the pretrained weights, code, evaluation framework, visualization tools, and tutorials. We anticipate that CryoDINO will enable increasingly generalizable computational pipelines for structural biology and accelerate the adoption of foundation models for cryo-ET.
%


{\small
\bibliographystyle{splncs04}
\bibliography{ref_2}
}
\end{document}


\section*{Supplementary Material}
\setcounter{figure}{0}
\setcounter{table}{0}
\renewcommand{\thefigure}{S\arabic{figure}}
\renewcommand{\thetable}{S\arabic{table}}
\renewcommand{\theHfigure}{S\arabic{figure}}
\renewcommand{\theHtable}{S\arabic{table}}

\begin{table}[!hbp]
\centering
\caption{\textbf{Dataset information for downstream segmentation.} For DS-10001, eight tomograms were used for training, and the same two tomograms were used for both validation and testing. All other datasets used separate training, validation, and test sets.}
\label{tab:tab_suppl1}
\resizebox{\textwidth}{!}{%
\begin{tabular}{lllll}
\hline
Dataset IDs  & \# tomograms & Classes & Species & Data split \\ \hline
DS-10001     & 10 & cytoplasm, organelles, membrane & Schizosaccharomyces pombe &
\begin{tabular}[c]{@{}l@{}}Train: 8\\ Val: 2\\ Test: 2 \end{tabular} \\ \hline
DS-10010     & 30 & membrane & Mus musculus &
\begin{tabular}[c]{@{}l@{}}Train: 21\\ Val: 4\\ Test: 5\end{tabular} \\ \hline
EMPIAR-10989 & 3 & actin & Homo sapiens &
\begin{tabular}[c]{@{}l@{}}Train: 1\\ Val: 1\\ Test: 1\end{tabular} \\ \hline
EMPIAR-12049 & 12 & membranes, microtubules, ribosomes & Caenorhabditis elegans &
\begin{tabular}[c]{@{}l@{}}Train: 8\\ Val: 2\\ Test: 2\end{tabular} \\ \hline
\end{tabular}%
}
\end{table}

\begin{table}[!hbp]
\centering
\caption{Dataset information for the downstream particle detection tasks}
\label{tab:tab_suppl2}
\resizebox{\textwidth}{!}{%
\begin{tabular}{lllll}
\hline
Dataset IDs  & \# tomograms & Classes & Species & Data split \\ \hline
\begin{tabular}[c]{@{}l@{}}DS-10440 (train)\\ DS-10445 (val)\\ DS-10446 (test)\end{tabular} &
492 &
\begin{tabular}[c]{@{}l@{}}apo-ferritin, beta-amylase, beta-galactosidase,\\ cytosolic ribosome, thyroglobulin, virus-like particle\end{tabular} &
In vitro mixture of purified complexes &
\begin{tabular}[c]{@{}l@{}}Train: 7 \\ Val: 121\\ Test: 364\end{tabular} \\ \hline 
BYU (Kaggle) & 362 & bacterial flagellar motor & Multiple species of bacteria included &
\begin{tabular}[c]{@{}l@{}}Train: 253 \\ Val: 54\\ Test: 55\end{tabular} \\ \hline 
\end{tabular}%
}
\end{table}

\begin{figure}[t]
  \centering
  \includegraphics[width=\textwidth]{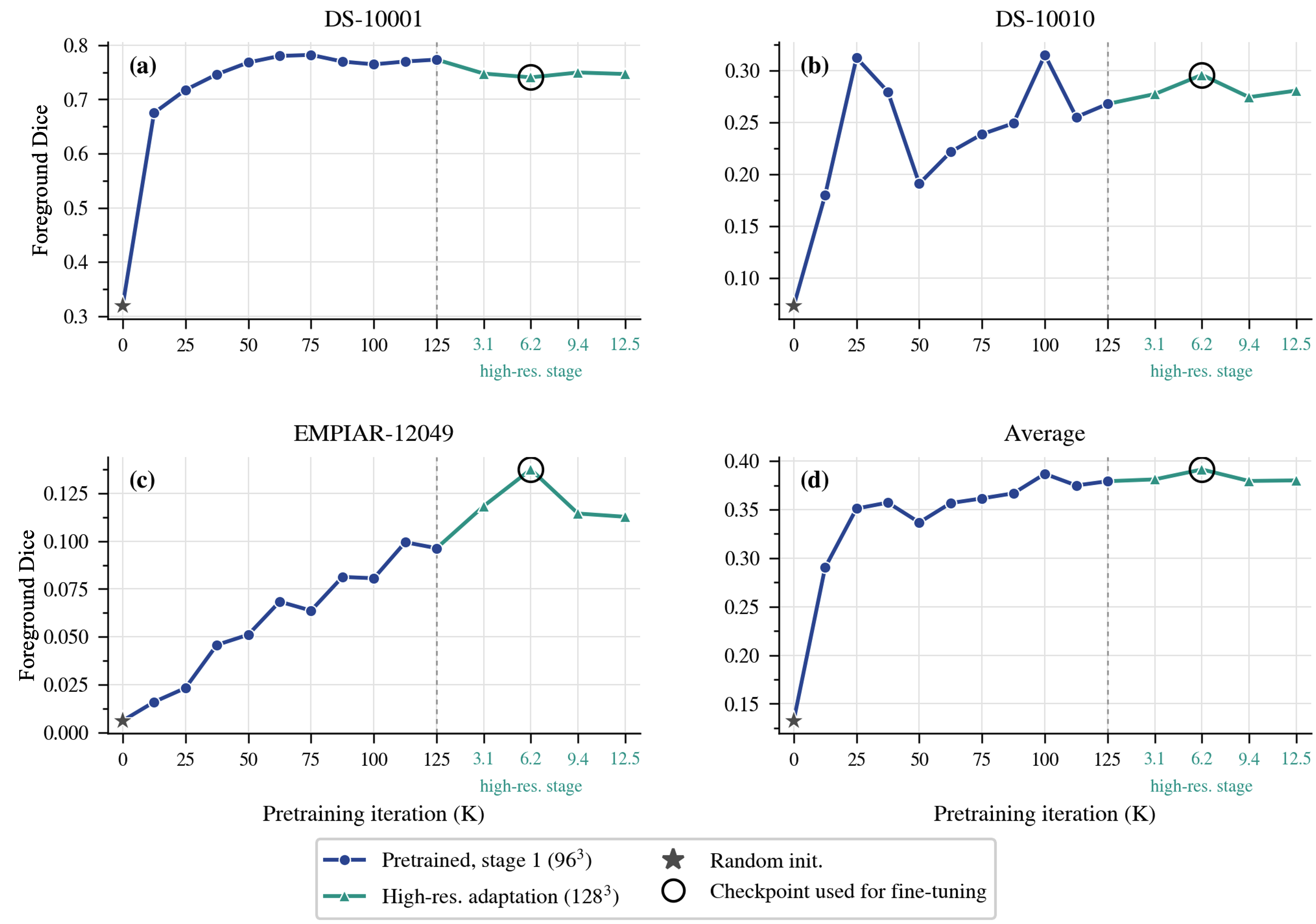}
  \caption{
  \textbf{Effect of self-supervised pretraining iterations on downstream segmentation performance.}
  Foreground validation Dice as a function of pretraining iterations for (a) DS-10001, (b) DS-10010, (c) EMPIAR-12049, and (d) their average. Stars mark random initialization, and open circles mark the checkpoint used for fine-tuning.
  A linear segmentation head is used for this experiment.
  }
  \label{fig:suppl1}
\end{figure}

\begin{figure}[t]
  \centering
  \includegraphics[width=\textwidth]{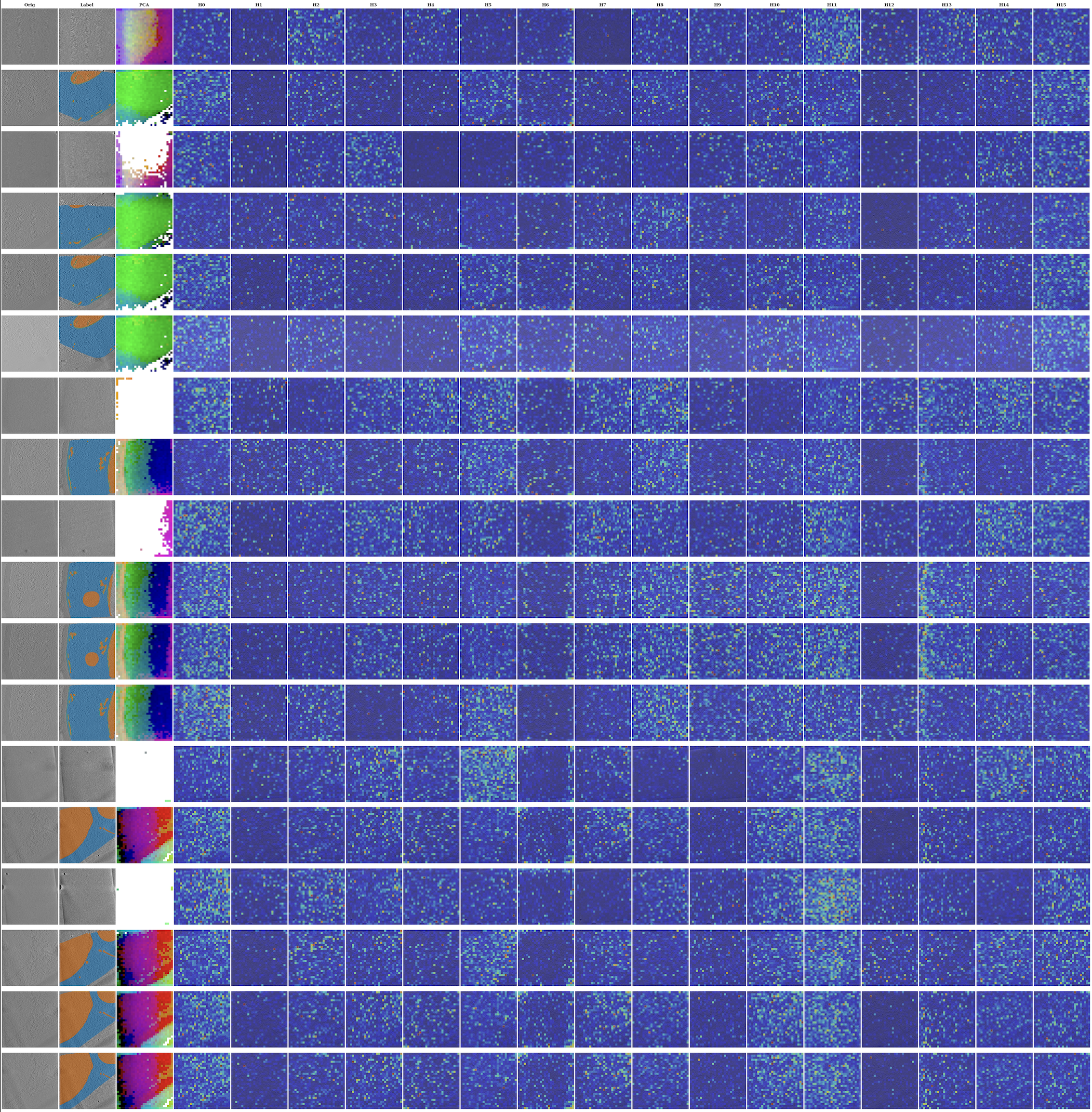}
  \caption{
  Original tomograms, segmentation labels, PCA visualization of patch-level representations, and multi-head self-attention maps from the 16 attention heads.
  }
  \label{fig:suppl2}
\end{figure}

\begin{figure}[p]
\centering
\includegraphics[width=\textwidth,height=0.80\textheight,keepaspectratio]{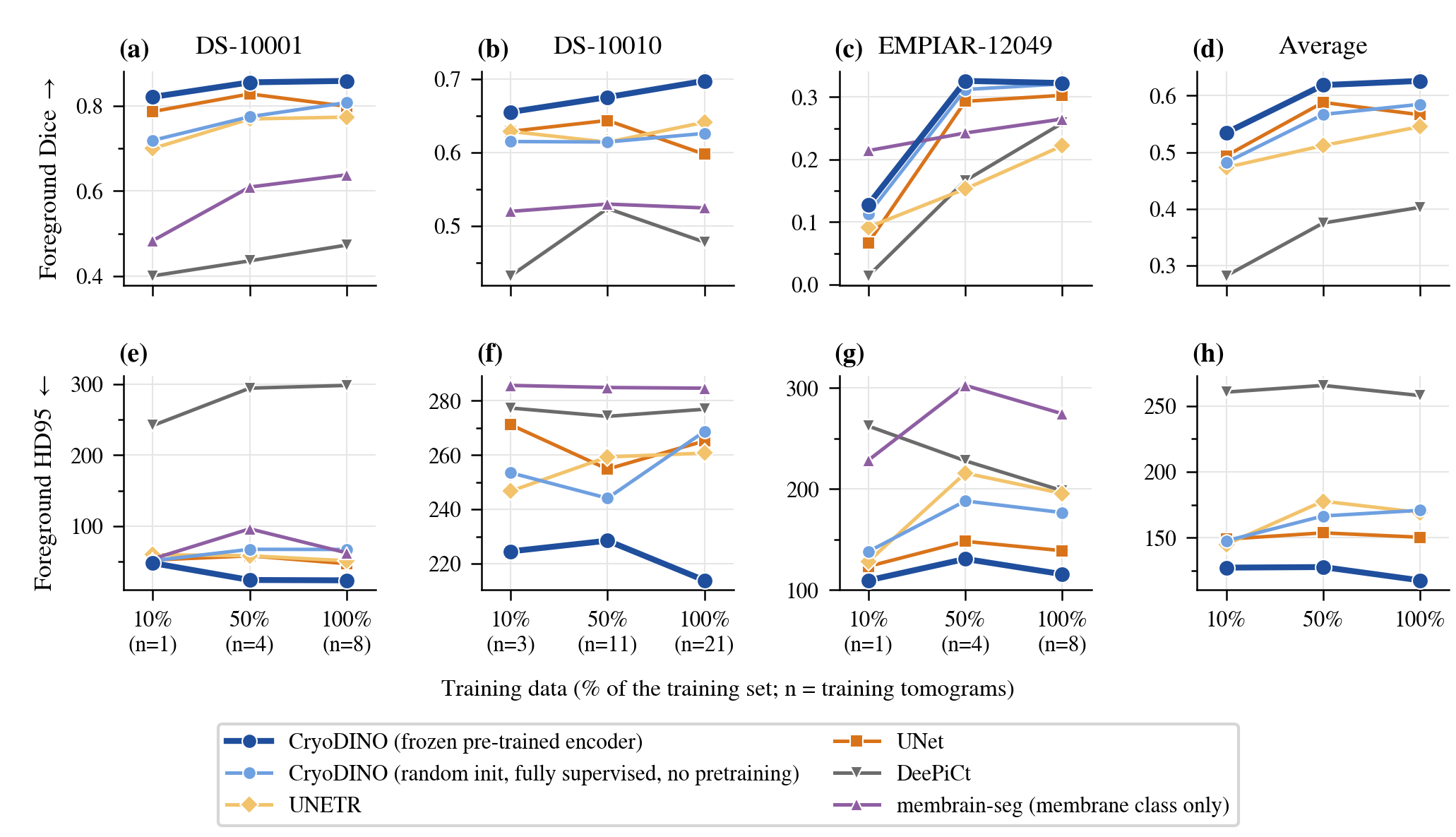}
\caption{\textbf{Segmentation performance across training-data proportions.} Foreground Dice (a--d) and HD95 (e--h) across labeled-data proportions for DS-10001, DS-10010, EMPIAR-12049, and the average across datasets. CryoDINO uses a frozen pretrained encoder.}
\label{fig:suppl3}
\end{figure}

%


\begin{figure}[t]
  \centering
  \includegraphics[width=0.74\textwidth]{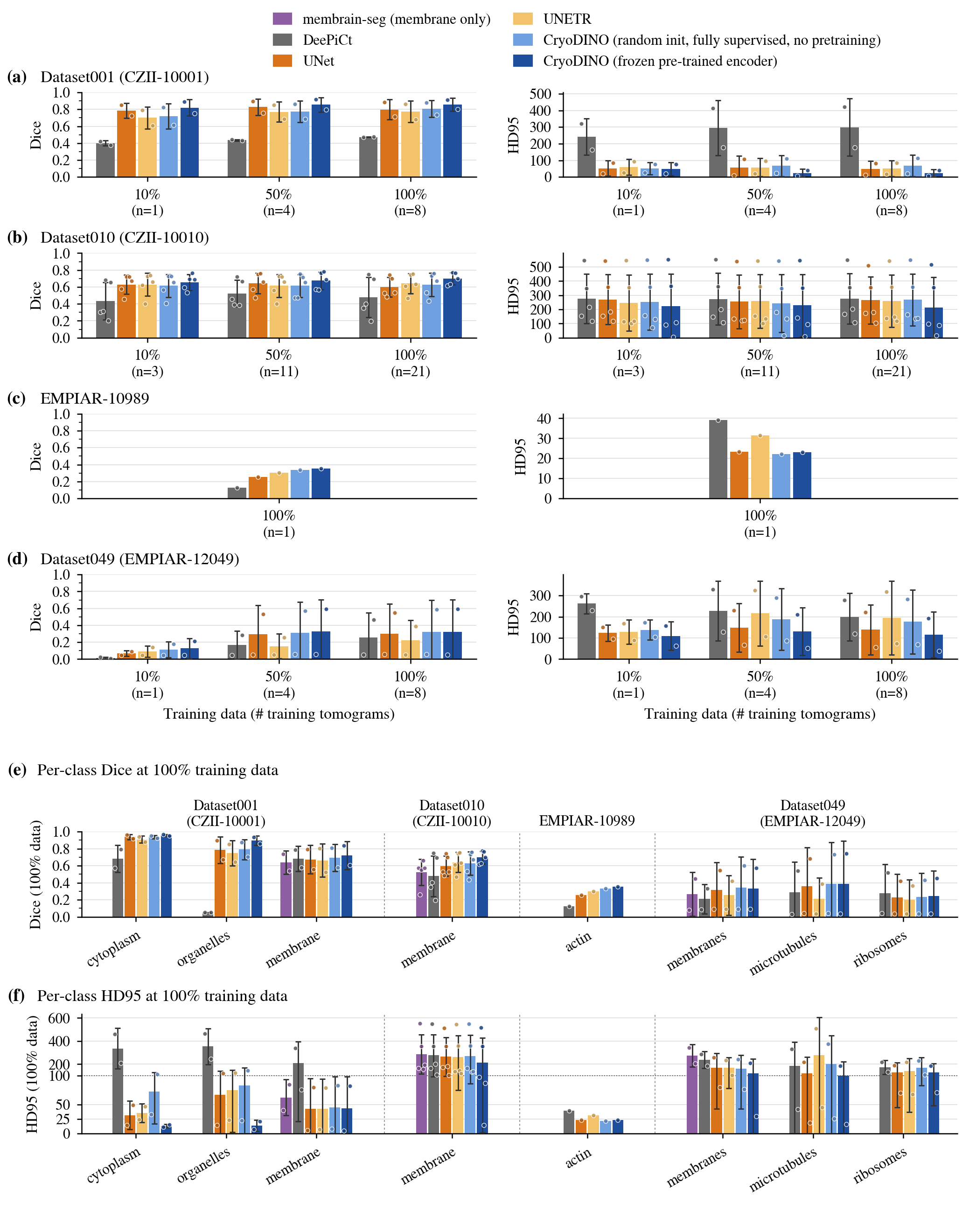}

\caption{\textbf{Quantitative segmentation performance.} (a--d) Foreground Dice and HD95 across four datasets and varying amounts of labeled training data. Bars show test-set means, dots indicate individual tomograms, and error bars denote $\pm1$ s.d. (e, f) Per-class Dice and HD95 using 100\% of the training data. CryoDINO, with a frozen pretrained backbone, consistently outperforms state-of-the-art methods across all four datasets.}
  \label{fig:fig2}
\end{figure}

\begin{figure}[p]
\centering
\includegraphics[width=\textwidth,height=0.80\textheight,keepaspectratio]{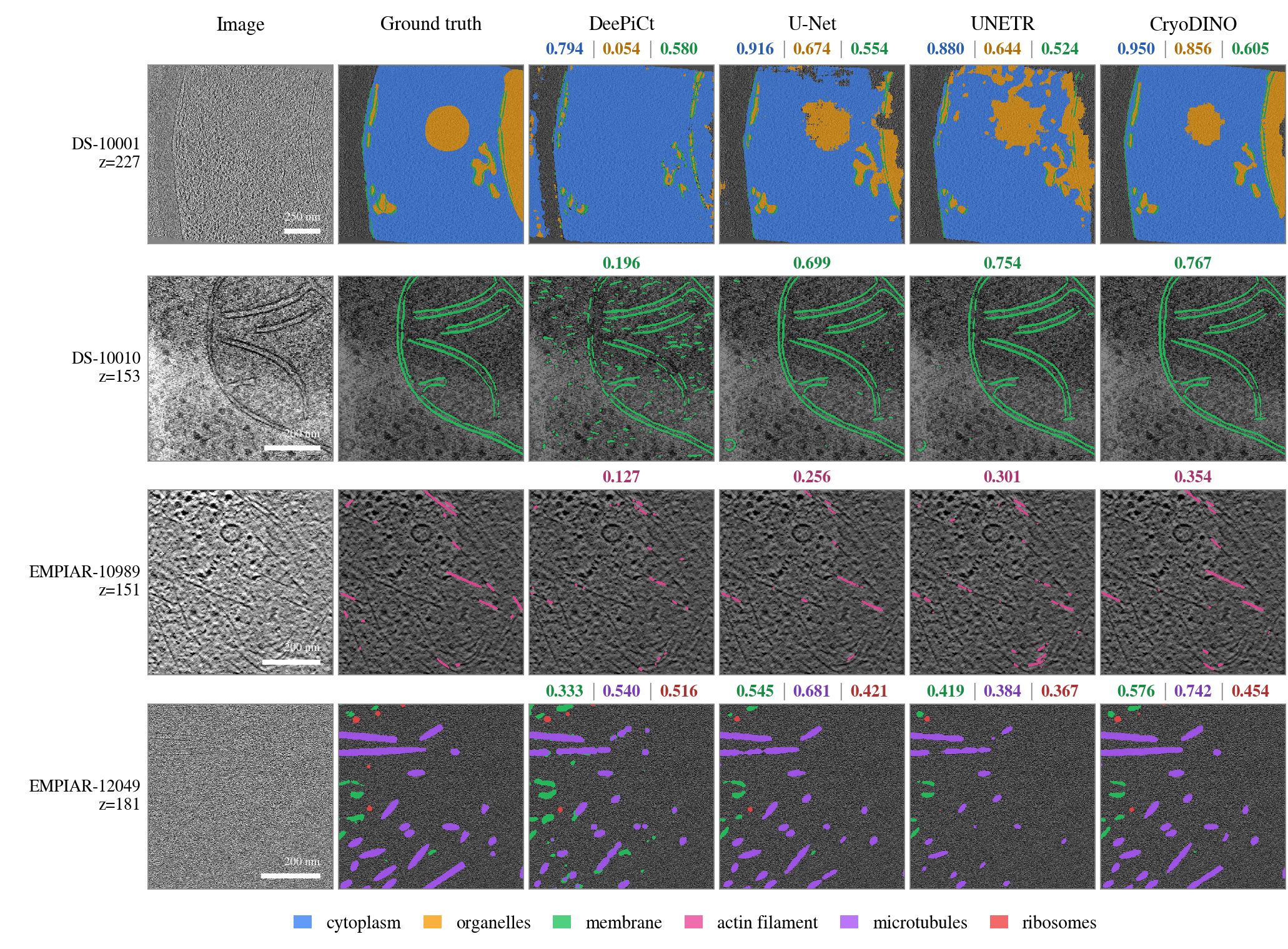}
\caption{\textbf{Slice-level semantic segmentation.} Representative slices from all four segmentation datasets, with ground truth and predictions from DeePiCt, U-Net, UNETR, and CryoDINO. Colors identify semantic classes, and numbers above predictions indicate per-class Dice scores.}
\label{fig:supp_slices}
\end{figure}

\begin{figure}[p]
\centering
\includegraphics[width=\textwidth,height=0.80\textheight,keepaspectratio]{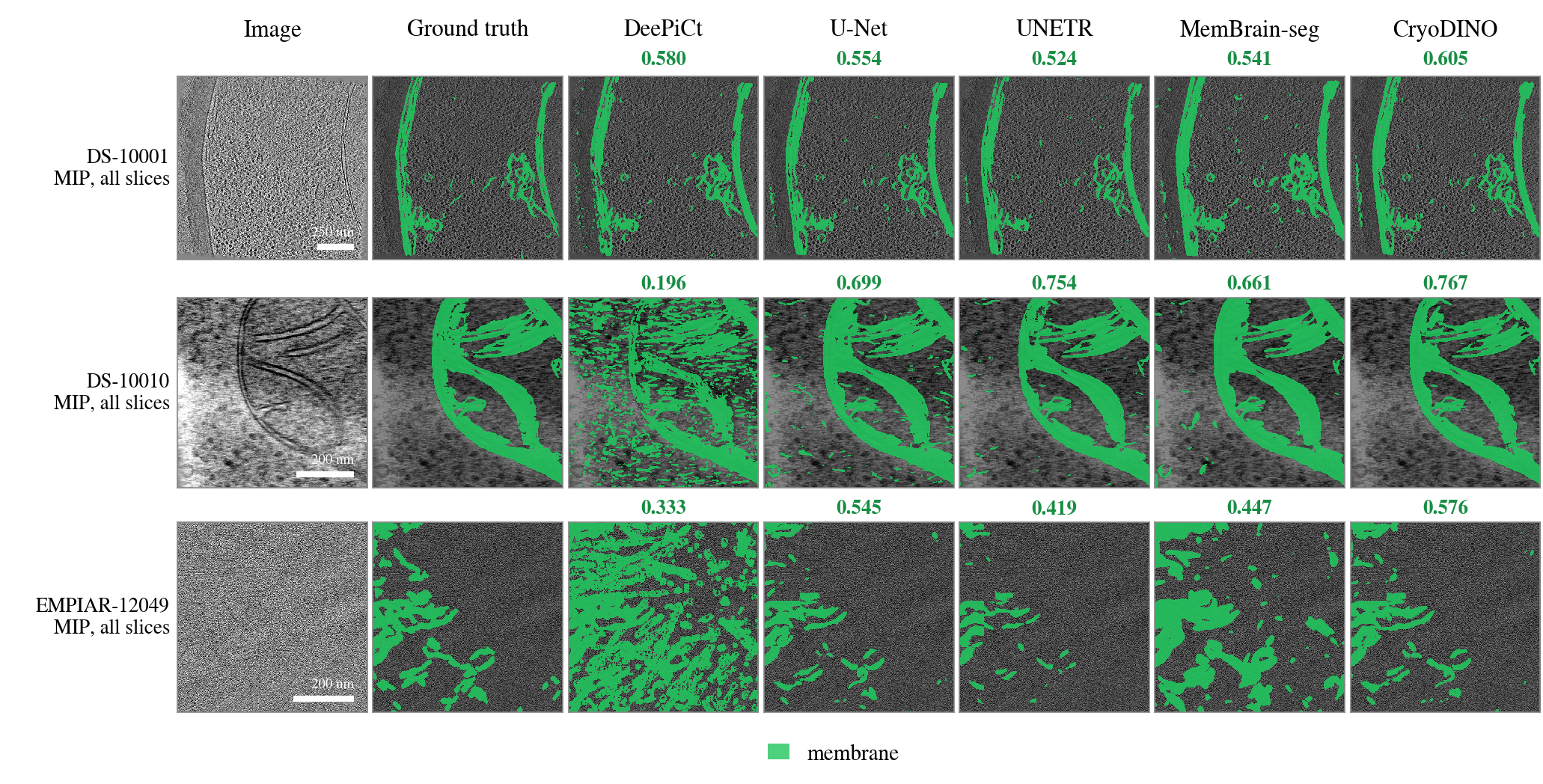}
\caption{\textbf{Membrane segmentation maximum-intensity projections.} Projections across all slices for DS-10001, DS-10010, and EMPIAR-12049, with ground truth and model predictions. Membranes are shown in green; numbers above predictions indicate Dice scores.}
\label{fig:supp_membranemip}
\end{figure}

\begin{figure}[p]
\centering
\includegraphics[width=\textwidth,height=0.80\textheight,keepaspectratio]{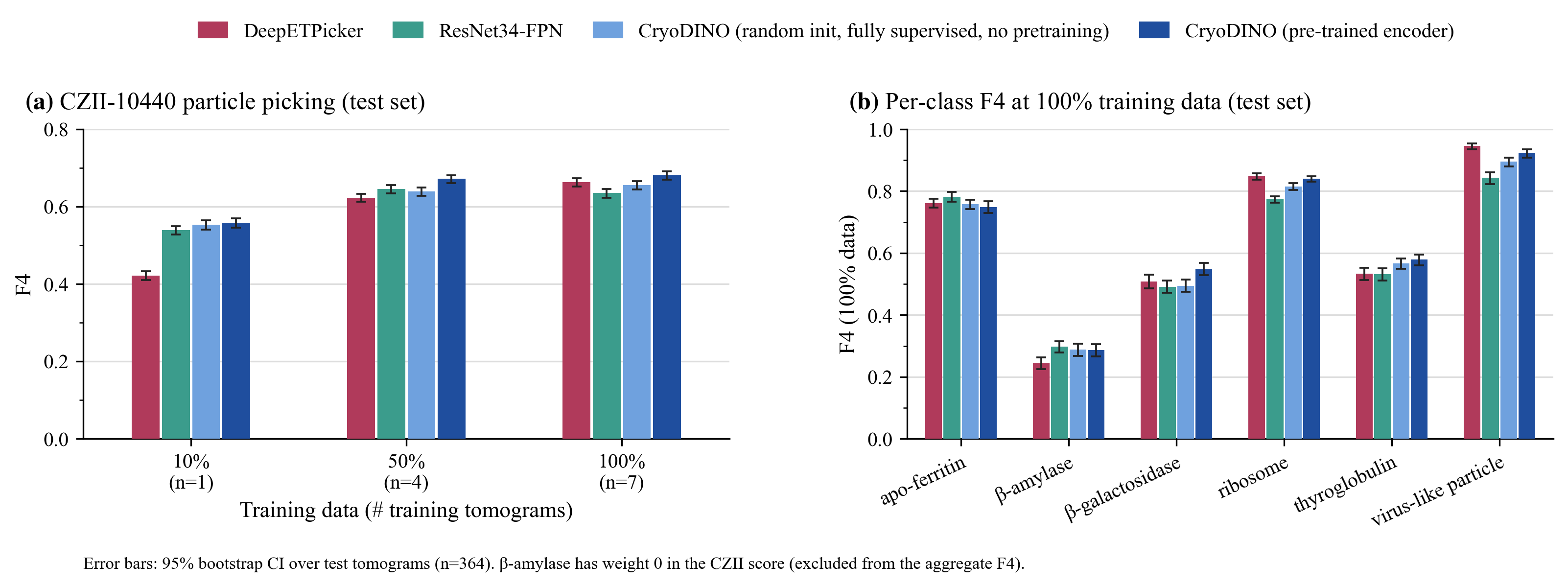}
\caption{\textbf{Particle-picking performance on the CZII dataset.} (a) Test-set aggregate F4 with 10\% (1 tomogram), 50\% (4 tomograms), and 100\% (7 tomograms) of the training data. (b) Per-class test-set F4 using all training data. Results compare DeepETPicker, ResNet34-FPN, CryoDINO trained from random initialization, and CryoDINO with a pretrained encoder. Error bars indicate 95\% bootstrap confidence intervals over the test tomograms. $\beta$-amylase has zero weight in the CZII score and is excluded from aggregate F4.}
\label{fig:s7}
\end{figure}

\begin{figure}[p]
\centering
\includegraphics[width=\textwidth,height=0.80\textheight,keepaspectratio]{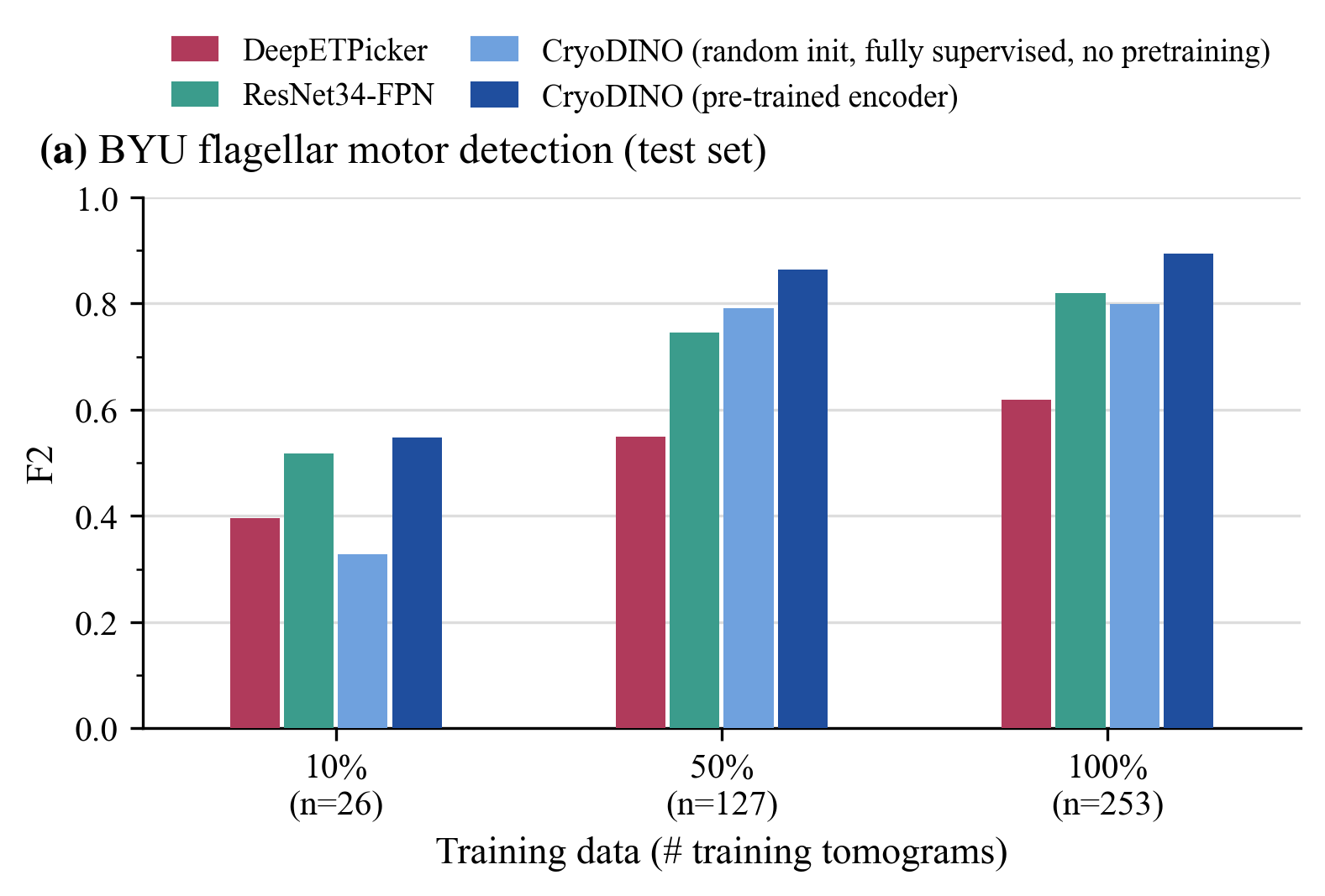}
\caption{\textbf{Particle-detection performance on the BYU bacterial flagellar motor dataset.} Test-set $F_2$ across training-data proportions (10\%, 50\%, and 100\%). Results compare DeepETPicker, ResNet34-FPN, CryoDINO trained from random initialization, and CryoDINO with a pretrained encoder.}
\label{fig:s8}
\end{figure}

\begin{figure}[p]
\centering
\includegraphics[width=\textwidth,height=0.80\textheight,keepaspectratio]{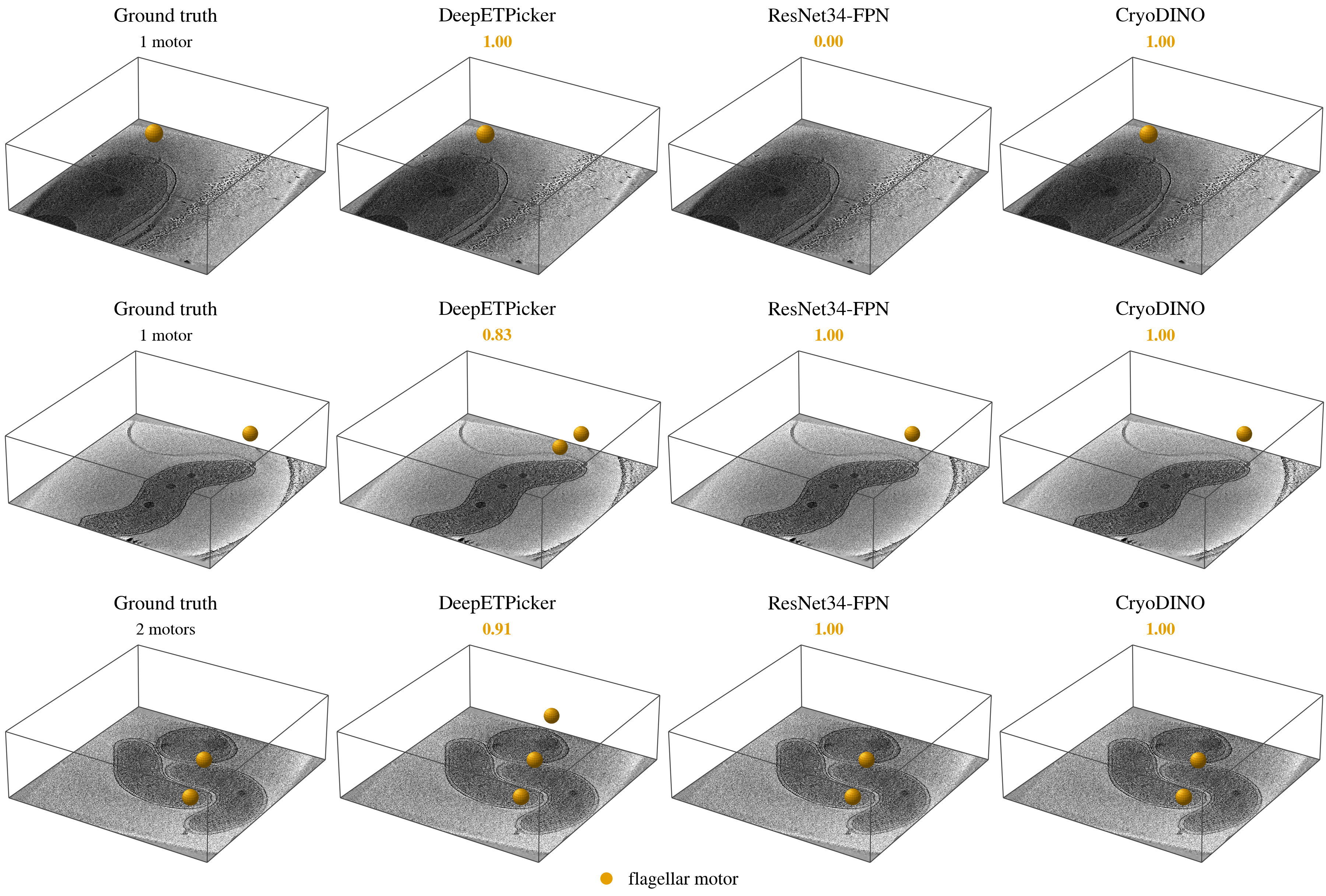}
\caption{\textbf{Qualitative particle-detection examples on the BYU bacterial flagellar motor dataset.} Three representative tomograms, showing ground-truth locations and predictions from DeepETPicker, ResNet34-FPN, and CryoDINO. Numbers indicate the F2 score for each example.}
\label{fig:s9}
\end{figure}